# Robust Protection of III-V Nanowires in Water Splitting by a Thin Compact TiO$_2$ Layer

Fan Cui[1†], Yunyan Zhang[1,8†*], H. Aruni Fonseka[2], Premrudee Promdet[3], Ali Imran Channa[4], Mingqing Wang[5], Xueming Xia[3], Sanjayan Sathasivam[3], Hezhuang Liu[4], Ivan P. Parkin[3], Hui Yang[6], Ting Li[7], Kwang-Leong Choy[5*], Jiang Wu[4*], Chris Blackman[3], Ana M. Sanchez[2] & Huiyun Liu[1]

1. Department of Electronic and Electrical Engineering, University College London, London WC1E 7JE, UK

2. Department of Physics, University of Warwick, Coventry CV4 7AL, United Kingdom

3. Department of Chemistry, University College London, London WC1H 0AJ, UK

4. Institute of Fundamental and Frontier Sciences, University of Electronic Science and Technology of China, Chengdu 610054, P. R. China

5. UCL Institute for Materials Discovery, University College London, Roberts Building, Malet Place, London, WC1E 7JE, UK.

6. Department of Materials, Imperial College London, Exhibition Road, London SW7 2AZ, UK

7. Institute of Biomedical Engineering, Chinese Academy of Medical Sciences & Peking Union Medical College, Tianjin, 300192

8. Department of Physics, Paderborn University, Warburger Straße 100, 33098, Paderborn, Germany

† These authors contributed equally to this work.

*E-mail: yunyan.zhang.11@ucl.ac.uk, k.choy@ucl.ac.uk, jiangwu@uestc.edu.cn

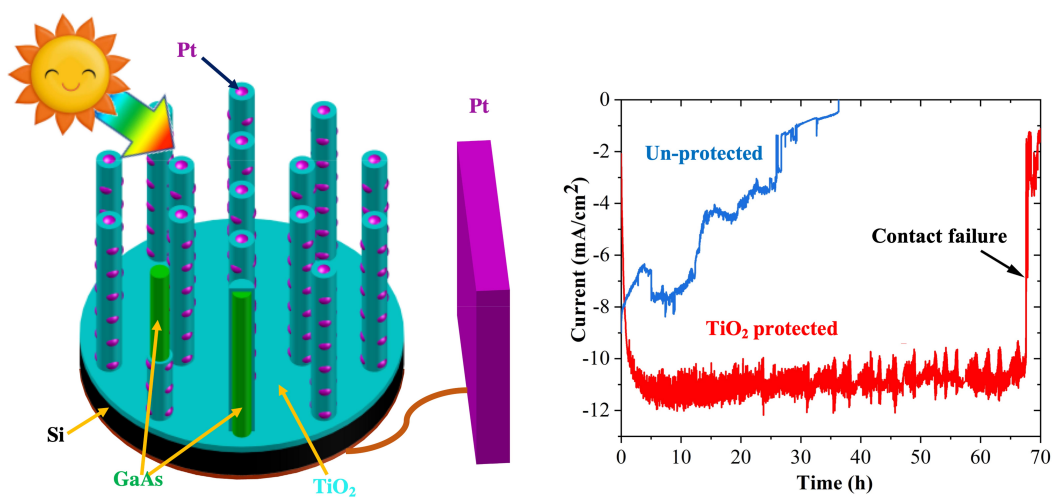

**Abstract:** Narrow-bandgap III-V semiconductor nanowires (NWs) with a suitable band structure and strong light-trapping ability are ideal for high-efficiency low-cost solar water-splitting systems. However, due to their nanoscale dimension, they suffer more severe corrosion by the electrolyte solution than the thin-film counterparts. Thus, short-term durability is the major obstacle for using these NWs for practical water splitting applications. Here, we demonstrated for the first time that a thin layer (~7 nm thick) of compact $TiO_2$ deposited by atomic layer deposition can provide robust protection to III-V NWs. The protected GaAs NWs maintain 91.4% of its photoluminescence intensity after 14 months of storage in ambient atmosphere, which suggests the $TiO_2$ layer is pinhole-free. Working as a photocathode for water splitting, they exhibited a 45% larger photocurrent density compared with un-protected counterparts and a high Faraday efficiency of 91%, and can also maintain a record-long highly-stable performance among narrow-bandgap III-V NW photoelectrodes; after 67 hours photoelectrochemical stability test reaction in strong acid electrolyte solution (pH = 1), they show no apparent indication of corrosion, which is in stark contrast to the un-protected NWs that are fully failed after 35-hours. These findings provide an effective way to enhance both stability and performance of III-V NW based photoelectrodes, which are highly important for practical applications in solar-energy-based water splitting systems.

**Introduction**

Solar energy is abundant, clean, and renewable, which makes it one of the most promising renewable energy sources that can solve the worldwide energy crisis and the serious environmental problems caused by the combustion of fossil fuels [1]. Solar-driven water-splitting can harvest solar energy and directly convert it into chemical energy, such as hydrogen from water reduction [2]. Dihydrogen has high energy density, which is beneficial for energy storage and transportation, and is particularly attractive as an energy carrier [3-5]. The combustion of hydrogen to water does not cause any environmental pollution. Hydrogen generation by photoelectrochemical (PEC) water splitting has thus gained great attention [4,6–9].

PEC water splitting requires a thermodynamic potential of ~1.23 eV and ~0.95 eV for full and Z-scheme half water splitting, respectively [10–12]. A potential much larger than these values will lead to low energy conversion efficiency due to energy loss [10]. For example, the ZnO with a large band gap of 3.37 eV can only covert a very small portion of high-energy photons in the solar spectrum. Thus, the high-efficiency semiconductor photo-electrode needs to have a narrow bandgap as close as possible to these ranges to convert a larger portion of energy from the solar spectrum [13–16]. The bandgap in many III-V semiconducting materials lie in this region.

Their direct bandgap allows high-efficiency photon absorption, making them suitable for high efficiency solar water splitting (~20%) [17]. However, III-V materials are relatively rare and expensive for extensive use. Innovations are needed to harvest solar energy with greater economic viability. The ideal solution is to build the high-efficiency III–V cells onto the low-cost mature Si platform. However, after 20 years of research, the lattice and thermal expansion coefficient mismatches between III–V epi-layers and Si substrates are still hindering the effective implementation of this idea [18].

III-V nanowire (NW) structures have demonstrated many novel mechanical, optical and electronic properties that are not present in the thin-film counterparts [19,20]. High-quality NWs can be grown on inexpensive substrates such as Si, graphene, carbon nanotubes, fibre-textured silicon thin film, amorphous Si, glass, and indium tin oxide, to significantly reduce the overall device cost [21]. They can also behave as optical antennas [22] to concentrate light as they can greatly enhance the light absorption cross-section (up to 12 times) compared to their physical size [23]. NW arrays can also increase light-scattering due to their subwavelength dimensions [24], resulting in the internal light path lengths up to 73 times longer compared with that of their thin-film counterparts [25]. Therefore, NW arrays have advanced light trapping ability, allowing to use a small amount of expensive III-V material and achieve as efficient optical absorption as thick bulk counterparts [26]. Moreover, the large surface-to-volume ratio of NWs provide lower the barrier for chemical reaction [4]. Thus, III-V NW based photoelectrodes for water splitting have attracted great attention.[27,28]

Severe photo-induced corrosion of III-V materials in the electrolyte solution is a common problem, and III-V NWs with a nano-scale size are even more vulnerable [29-31]. Due to the unique 1-dimensional structure, they are much more difficult to protect. The lifetime of the unprotected NW water splitting cells is commonly less than 1 day that is far too short to have practical applications [31]. Titanium dioxide ($TiO_2$) is one of the most common protection layer materials that is stable over a wide range of pH and potentials [18,32]. It also forms a type-II heterojunction with most III-V materials and serves as an effective hole blocking layer while allowing the transport of electrons to the surface for water splitting [10,33]. This allows it to reduce the carrier loss by non-radiative surface recombination, and effectively improve the quantum efficiency [32]. Thus, it has been widely used in III-V thin-film devices, and has demonstrated quite robust protection against electrolyte corrosion [34]. When grown by atomic layer deposition (ALD), it can conformably cover 3-dimensional sample surface with a uniform and precise thickness, therefore giving good protection to III-V NW during the water splitting process. Despite some initial studies [33], there is still a lack of reports on the achievement of a long-term stability when directly using $TiO_2$ protection for III-V NWs for water splitting. The thick $TiO_2$ layers of tens of nanometres or even hundreds of nanometres that have been used to achieve long lifetime protection, lead to parasitic light absorption and poor charge transfer characteristics [34,35]. Moreover, the increase in $TiO_2$ thickness can also increase the overpotential with a linear rate of ~21 mV/nm, which results in an additional voltage loss [36,37]. Therefore, it is important to develop a robust protection technique using a thin (sub 10 nm) layers of $TiO_2$ [38,39].

In this study, the protective behaviour of a very thin $TiO_2$ layer on III-V NWs was studied in depth. With $TiO_2$ protection, the GaAs NW photocathode demonstrated greatly improved performance and a record long durability in the photoelectrochemical reaction among narrow-bandgap III-V NWs. This solves a major challenge for using narrow-bandgap III-V NWs in the water splitting for environmentally clean and renewable energy generation.

**Results and discussion**

**Structural Information**

GaAs NWs were grown by molecular beam epitaxy (MBE) via self-catalysed mode on p-type Si substrates [40]. The majority of NWs are standing vertically on the substrate as can be seen in the scanning electron microscope (SEM) image in Figure 1a. They are ~10 μm in length with a diameter of ~ 300 nm at the bottom, which gradually increase to ~550 nm at tip. The lower two thirds of the NW is zinc blende crystal structure with occasional single twins, as can be seen in the transmission electron microscopy (TEM) image in Figure 1b. The NW tip is enlarged with an irregular shape caused by un-optimized consumption of the Ga catalytic droplet after the core-NW growth. Defect-free core-shell NWs with high crystalline quality and regular morphology can be achieved by optimized growth parameters [41].

NWs were then coated by an uniform layer of amorphous $TiO_2$ deposited by ALD, confirmed by the energy dispersive X-ray spectroscopy (EDX) composition mapping of Ga and Ti shown in Figure 1c and the composition line scan near the surface shown in Figure 1d, as well as the X-ray photoelectron spectroscopy (XPS) spectra in Supporting Information Figure S1a. The thickness of $TiO_2$ protection layer is ~7 nm with a high uniformity from the bottom to the tip of the NWs, as shown in Figure 1e~h. The uniform deposition of conformal films with highly controllable thickness on complex three-dimensional surfaces is a key benefit of ALD growth. The Pt co-catalyst was deposited on the surface of $TiO_2$ by aerosol assisted chemical vapor deposition (AACVD). Pt formed small particles (<3 nm) which are confirmed by EDX line scans in Figure 1d. The Pt nanoparticles are decorating the surface of the NWs as shown in Figure 1h~k, but the density and size gradually reduce from the tip to bottom, possibly due to the shadowing effect caused by the neighbouring NWs.

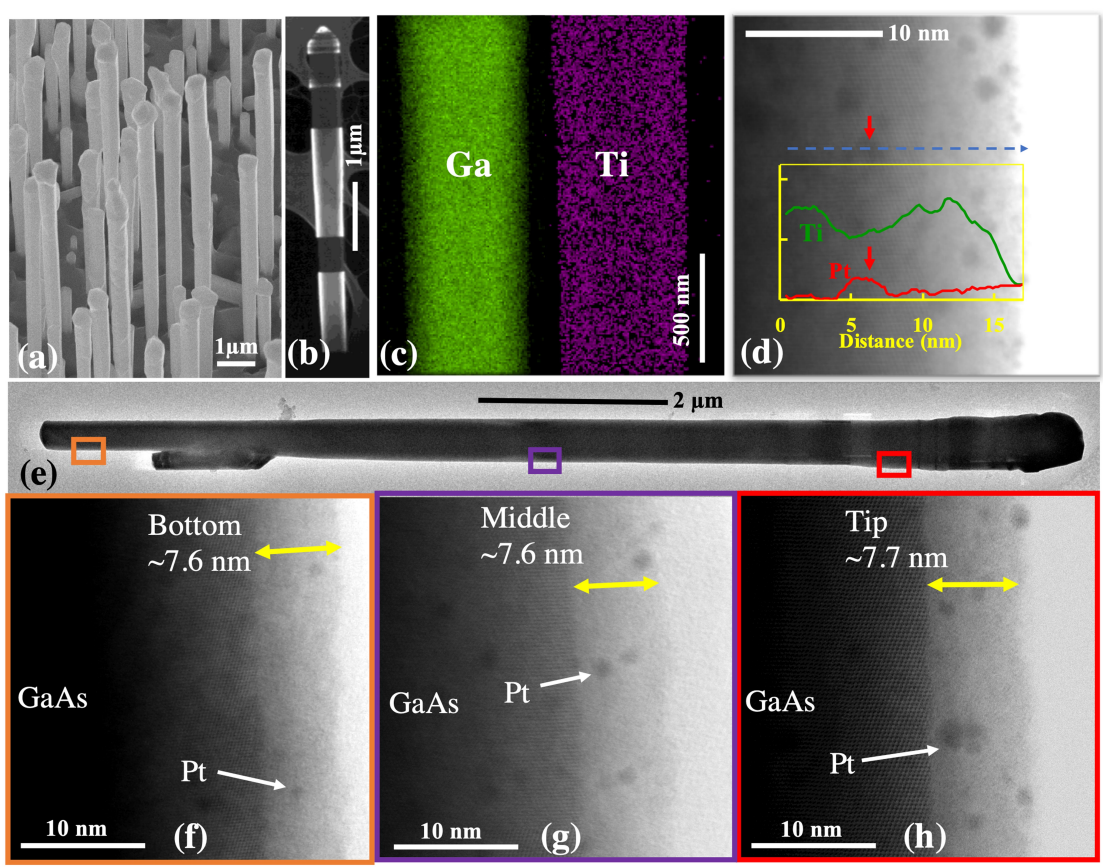

**Figure 1**. Structure information of GaAs NWs covered by $TiO_2$ and Pt. (**a**) Low-magnification SEM images showing the NWs. (**b**) Low-magnification TEM image of a NW. (**c**) EDX mapping of Ga and Ti in a NW segment. (**d**) EDX composition line scan near the NW surface shown in its inset. (**e**) Low-magnification TEM image showing an entire NW. Higher

magnification TEM images showing the NW/TiO$_2$ interfaces at the (**f**) bottom, (**g**) middle, and (**h**) tip of the NW marked in (e).

## Compact TiO$_2$ and its influence on optical properties

The influence of TiO$_2$ on the optical properties of GaAs NWs was analysed by photoluminescence (PL) spectroscopy. Figure 2 shows the PL spectrum of GaAs NWs with and without the TiO$_2$ layer. With the addition of TiO$_2$ on the same day after the NW growth, PL emissions from GaAs NWs are quenched by a factor of 2.8, which is typical for type-II heterojunctions that can efficiently separate charge carriers and thereby reduce radiative recombination [10]. As illustrated in the inset of Figure 2, the conduction band (CB) of GaAs is slightly higher than that of TiO$_2$, therefore electron migration from the NWs to the surface is promoted allowing for efficient water splitting. The valence band of TiO$_2$ is much lower than that of GaAs, thus forming a favourable barrier to keep holes away from the surface.

When exposed to air, the NW surface can be oxidised to form a native oxide layer (1-2 nm), which act as high-density non-radiative recombination centres, resulting in charge trapping. This can consume a large portion of the photon generated carriers and lead to low light emission. After storing in atmosphere for 14 months, the NWs without TiO$_2$ protection experienced severe decay and the emission intensity reduced to 13.2%. In contrast, the NWs with a ~7 nm TiO$_2$ protection layer can maintain 91.4% of the emission intensity after storing in the same environment and period. This suggests that the TiO$_2$ layer has good compactness that can prevent the permeation of oxygen and water, and thus can provide a long-term protection to the NWs' surface.

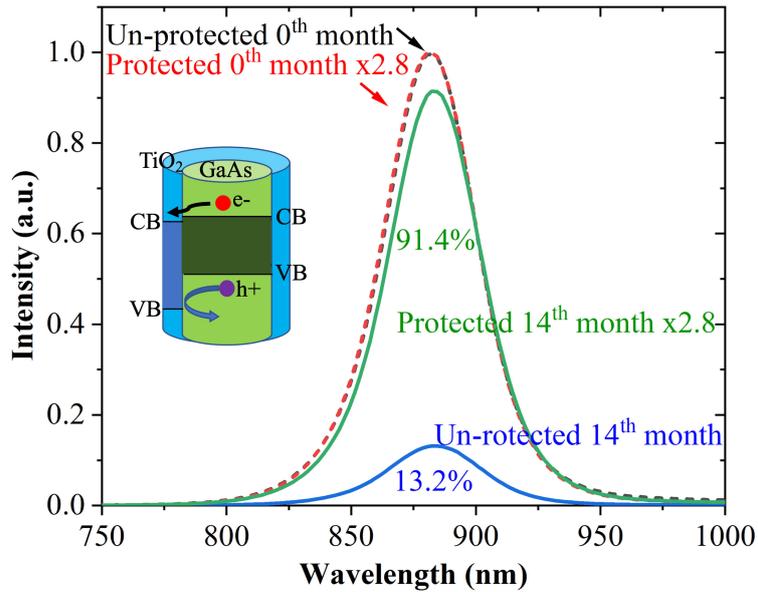

Figure 2. PL spectra from GaAs NWs with and without a $TiO_2$ protection layer. The $TiO_2$ protection layer was deposited on the same day of the NW growth. The spectra were taken on the 1st day of NW growth and after 14 months' storage in atmosphere.

**Photoelectrochemical Performance**

The PEC performance of GaAs NWs photoelectrodes with and without $TiO_2$ was determined by using them as a photocathode to water splitting and $H_2$ generation. A three-electrode system, working electrode, Ag/AgCl as reference electrode and Pt as counter electrode, was used. 0.5 M $H_2SO_4$ (pH = 1) solution was used as electrolyte, with the PEC reaction carried out under 1 sun illumination from AM 1.5 G solar simulator.

As shown in Figure 3a, the un-protected GaAs NW photocathodes have a photocurrent onset potential of ~0.233 V reversible hydrogen electrode (RHE), and a photocurrent density of 0.598 mA/cm² at 0 V vs RHE. As shown in Figure 3b, the protected GaAs NW photocathodes have a slightly larger photocurrent onset potential of ~0.243 V RHE, and a much larger photocurrent density of 0.87 mA/cm² at 0 V vs RHE – a 45% increase. The increased photocurrent density is due to $TiO_2$ surface passivation, and the type-II band alignment between $TiO_2$ and GaAs NWs that are beneficial for efficient carrier separation as discussed above.

A gas chromatograph (GC) system was used to measure the actual $H_2$ generation rate of the protected NW photocathode when under continuous irradiation with a bias voltage of -0.6 V. As shown in Figure 3c, the generated $H_2$ volume increased linearly. The theoretical generation value of hydrogen is obtained from the I×t curve by the equation [42]:

$$H_2 = \frac{Q}{2F} = \frac{I \times t}{2F} = \frac{1}{2}\frac{\left(\int_0^t I dt\right)}{F} \quad \dots\dots\dots(1)$$

Where, F is the Faraday constant (96484.34 C/mol), Q is the amount of charge passed in time t, I is the photocurrent. when the current is not constant, the amount of charge passed through the circuit can be estimated by integrating the current over time. The Faraday efficiency can be calculated by the equation:

$$\text{Faraday efficiency} = \frac{\text{Experimental generation value of } H_2}{\text{Theoretical estimated value of } H_2} \quad \dots\dots\dots(2)$$

As can be deduced from Figure 3c, the $H_2$ generation rate was about 22 μmol/h and is highly stable for the one-hour period measured, which can be further supported by the stable photocurrent in the inset. The photocathode also exhibited a high and stable Faraday efficiency of 91%, which shows that the surface condition is highly favourable for carriers to participate in the water-splitting reaction.

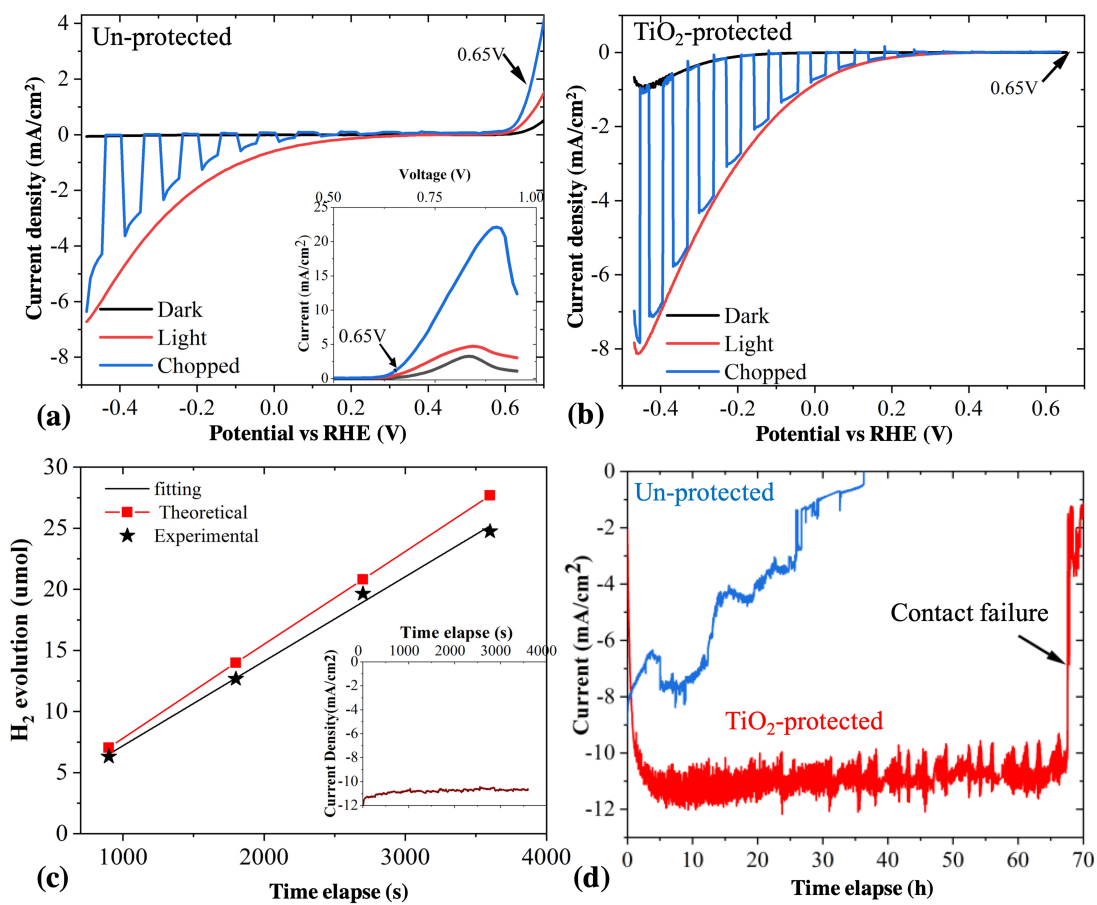

**Figure 3**. Photoelectrochemical properties of GaAs photocathodes. (**a**) The J-V curves of un-protected photocathodes and the inset Figure. shows the enlarged of J-V curve above 0.5V vs RHE. (**b**) $TiO_2$-protected photocathodes under one sun illumination (AM 1.5 G 100 mW/cm$^2$) in 0.5 M $H_2SO_4$ electrolyte (pH=1). (**c**) Experimental and theoretical $H_2$ evolution of $TiO_2$-protected photocathodes under continuous illumination (AM 1.5G 100 mW/cm$^2$) (**d**) Stability of the protected and un-protected photocathodes measured with a bias of -0.6 V and under an AM 1.5 sun illumination.

**Photoelectrochemical Stability**

The corrosion current of the un-protected NW photocathode is ~2 mA/cm$^2$ at 0.65 V vs RHE and increased rapidly with a forward bias as can be seen in Figure 3a and its inset; while the TiO$_2$-protected photocathodes do not show photo-corrosion current at 0.65 V vs RHE, which suggests better stability in the PEC reaction. To further confirm this, a stability test of the hydrogen evolution reaction of the photocathodes was carried out. The hydrogen is uniformly generated on the surface of the test samples, and the hydrogen generation rate is fast and stable when the NWs were in a good state (Supporting Information S2). As shown in Figure 3d, the current density of the un-protected NW photocathode started at about 8.5 mA/cm$^2$ under -0.6V vs RHE, then dropped rapidly to ~0 mA/cm$^2$ after ~36 hours; while the initial current density of the protected photocathode was ~11 mA/cm$^2$ and highly stable over a long duration of 67 hours that is already much longer than other narrow bandgap III-V NW photoelectrodes so far as we know. Beyond this time, no data is available due to the failure of the back contact (Supporting Information S3).

After the stability test, the NWs from the un-protected photocathode were found to have detached from the substrate and bundled together, forming large NW groups and producing areas free of NWs, as can be seen in Figure 4a and b. As can be seen in more detailed in Figure 4c corresponds to the bottom of the NWs, which is highly rough and significantly thinned down to ~50 nm due to the corrosion. NWs with a high-aspect ratio tend to bend and bundle together after removal from liquid due to the capillary force [43]. When the NW bottom is very thin, they can break off the substrate when bending. In addition, the other parts of NWs are also corroded during the PEC reaction, and the Pt particles seen to have been washed away in the process as can be seen in Figure 4d. For the protected photocathode, NWs are also bent and bundled, forming smaller groups of NWs as can be seen in Figure 4e. However, none of the NWs are observed to be broken off from the substrate due to solution etching. The TiO$_2$ protection layer is still on the surface of NWs with a thickness of ~7 nm (Figure 4f), which can be further confirmed by the uniform distribution of Ti from the EDS mapping shown in Figure 4g that were taken at the bottom of NWs. There is no noticeable changes in chemical format of TiO$_2$ films during PEC water splitting (Supporting Information S1). Besides, the Pt particles are also still present on the surface as can be seen in Figure 4f. Thus, no apparent corrosion from protected NWs was observed.

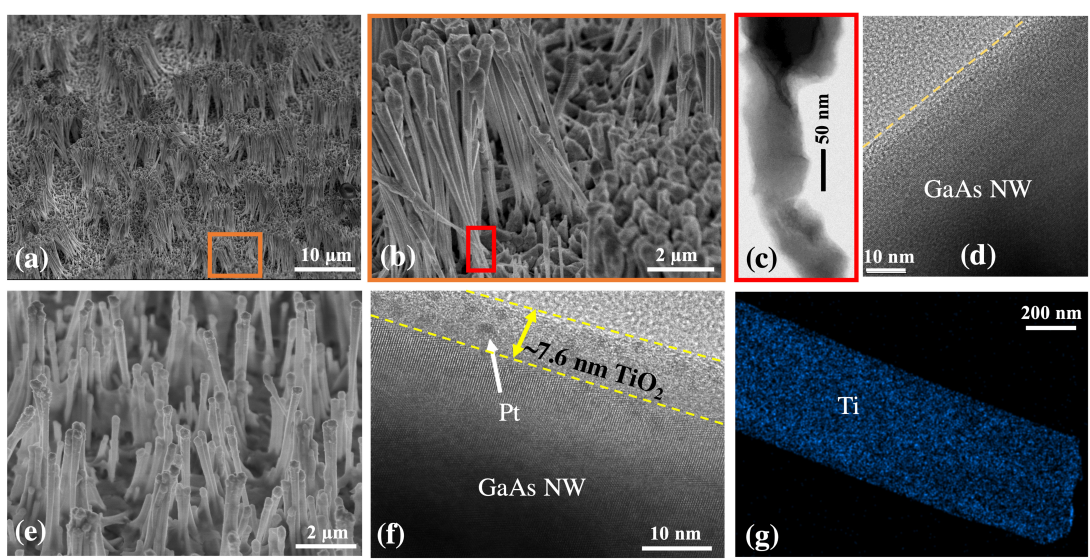

**Figure 4.** Morphology and structure of GaAs NW photocathodes after stability test. (a)~(d) are from un-protected photocathode. (**a**) SEM image gives an overall view of the sample surface. (**b**) higher magnification SEM image of the orange square shown in (a). (**c**) TEM image showing the bottom of a NW with severe corrosion. (**d**) TEM image showing the Pt-free surface of the NWs. (e)~(h) are from protected photocathodes. (**e**) SEM image of the sample surface.

(**f**) TEM image of NW surface showing GaAs/TiO$_2$ interface and the Pt particles. (**g**) is Ti element mapping of the NW bottom.

**Conclusion**

In this study, the stability of un-protected and protected narrow-bandgap III-V NWs in the PEC reaction was investigated in detail using GaAs NWs. The un-protected NWs suffered from severe corrosion, leading to full failure in the PEC reaction within 35 hours. TiO$_2$ layer of ~7 nm was used to protect the NWs and to minimize the detrimental corrosion effects. The ALD growth can conformably cover the entire NWs surface with a highly-uniform TiO$_2$ layer. After storing in atmosphere for 14 months, the protected NWs can still keep 91.4% of its emission intensity, outperforming the un-protected ones that can only maintain 13%. This suggests that the compact TiO$_2$ film can prevent oxygen and water permeation to the surface of GaAs NWs. The protected GaAs NWs were used as the cathode for PEC water splitting and the photocurrent density improved by 45% to 0.87 mA/cm$^2$ at 0V vs RHE due to the surface passivation effect from the TiO$_2$ layer and the favourable band alignment between GaAs and TiO$_2$, which improved the charge transfer and reduced the charge recombination at the photocathode. After 67 hours of PEC reaction under continuous simulated solar light illumination, there was no obvious decay of the electrode and the protection layer is still in a very good state. The protected photocathode also shows good faradaic efficiency of ~91%, which suggests that the surface condition is highly favourable for carriers to participate in the water-splitting reaction. These results show that a thin layer of compact TiO$_2$ can provide superior protection to GaAs NWs, and addresses for the first time the short stability issue of narrow bandgap III-V NWs and allows them to be practically used in the PEC reaction for significantly longer.

**Methods**

*NW growth:* The self-catalysed GaAs NWs were grown directly on p-type Si(111) substrates by solid-source III−V molecular beam epitaxy [44]. The Ga beam equivalent pressure, V/III flux ratio, and substrate temperature were $8.41\times10^{-8}$ Torr, ~50, and ~630 °C, respectively. To grow the shell, the Ga droplets were consumed by closing the Ga flux and keeping the group-V fluxes open after the growth of the core. The shells were grown with a Ga beam equivalent pressure, V/III flux ratio, substrate temperature, and growth duration of $8.41 \times 10^{-8}$ Torr, 86, ~500 °C, and 90 mins, respectively. The substrate temperature was measured by a pyrometer. The doping concentration of GaAs p-core, p-shell, and n-shell were $1.6\times 10^{18}$ (Be), $1.6\times 10^{18}$ (Be), and $1\times 10^{18}$~$1\times 10^{19}$ (Si) cm$^{-3}$, respectively. The thickness ratio of the GaAs p-core, p-shell, and n-shell is 2:3:3. On the p-i-n junction, a layer of ~30nm Al$_{0.5}$Ga$_{0.5}$As surface passivation layer and a ~10nm GaAs protection layer were grown with a Si doping concentration of $1\times 10^{19}$ cm$^{-3}$.

*Scanning Electron Microscope (SEM):* The NW morphology was characterised with a Zeiss XB 1540 FIB/SEM system.

*Transmission Electron Microscopy (TEM):* Simple scraping of the NWs onto a holey carbon support was used to prepare TEM specimens. The TEM measurements were performed with a doubly−corrected ARM200F microscope, operating at 200 kV.

***TiO₂ Deposition***: The uniform amorphous $TiO_2$ film was deposited on the GaAs nanowire at 100°C by applying ALD at the pressure of $1\times10^{-2}$ mbar, using titanium isopropoxide (TTIP) and water as metal and oxygen precursors [45]. The deposition rate was about 0.48Å/cycle.

***Pt Deposition***: Pt was applied as a co-catalyst deposited on the surface of GaAs nanowires coated with $TiO_2$ by the AACVD method to accelerate the hydrogen evolution reaction (HER) rate [46–48]. Aerosols of Pt precursor dissolved in methanol were generated by an ultrasonic humidifier and then the aerosols were transported to the reactor by using nitrogen carrier gas operated by a mass flow controller [49,50]. The deposition was carried out at 350°C with a deposition time 30 minutes.

***Photoelectrochemical Measurements (PEC):*** The PEC properties of the GaAs nanowire photoelectrodes were measured by a three-electrode system composed of a Ag/AgCl reference electrode, Pt counter electrode and GaAs NWs as the photoactive working electrode. The constant potential of the working electrode was controlled by a potentiostat. The electrolyte, 0.5M $H_2SO_4$ (pH = 1) was used as the PEC reaction solution, and the PEC reaction was carried out under 1 AM 1.5 G sun illumination. Stability test of the photocathode was carried out under an AM 1.5 sun illumination. The stability reaction was carried out in a 0.5 mol/L sulfuric acid solution (pH = 1), with a constant potential relative to an RHE equal to 0. A gas chromatograph (GC; Shimadzu GC-2014) was used to measure the hydrogen generation rate. According to the Nernst equation, the measured potential (vs. Ag/AgCl) can be converted into a reversible hydrogen electrode (NHE at pH = 1): $E_{RHE} = E_{Ag/AgCl} + E_{0Ag/AgCl} + 0.059 \times pH$. Generally, at room temperature $E_{0Ag/AgCl} = 0.197$ V [51]. In order to control photo corrosion, both the PEC test and the stability test voltage were selected as the safe voltage range -0.733–0.4 V. The photocurrent density (J) and electric potential (V) for a GaAs nanowire Si/GaAs-$TiO_2$-Pt photocathode was measured by linear scanning voltammetry. The measurement was carried out in the dark, in a chopper, and in continuous light under one sun illumination (AM 1.5 G 100 mW/cm$^2$) [52].

**Acknowledgements:**

The authors acknowledge the support of Leverhulme Trust, EPSRC (grant nos. EP/P000916/1, EP/P000886/1, EP/P006973/1), and EPSRC National Epitaxy Facility. The authors also acknowledge Research Interest limited company for capturing Figure 4f and g.

**Competing interests**

The authors declare no competing interests.

**Additional information**

Supplementary information is available for this paper at https://doi.org/